\documentclass{egpubl}
\usepackage{eurovis2024}

\SpecialIssuePaper         % uncomment for final version of Computer Graphics Forum, special issue
\AuthorPreprint

\usepackage[T1]{fontenc}
\usepackage{dfadobe}  

\usepackage{cite}  % comment out for biblatex with backend=biber

\BibtexOrBiblatex
\electronicVersion
\PrintedOrElectronic
\ifpdf \usepackage[pdftex]{graphicx} \pdfcompresslevel=9
\else \usepackage[dvips]{graphicx} \fi

\usepackage{egweblnk}

\usepackage{subcaption}
\usepackage{pifont}
\usepackage{bm}
\usepackage{amssymb}
\usepackage{enumitem}
\usepackage{xspace}
\usepackage{todonotes}

\newcommand{\supermarket}{supermarket\xspace}
\newcommand{\col}{\textit{Color}\xspace}
\newcommand{\arrow}{\textit{Arrow}\xspace}
\newcommand{\outline }{\textit{Outline}\xspace}
\newcommand{\link}{\textit{Link}\xspace}

\newcommand{\vsection}[1]{\vspace{-2mm}\section{#1}\vspace{-1mm}}
\newcommand{\vsubsection}[1]{\vspace{-2mm}\subsection{#1}\vspace{-1mm}}
\newcommand{\vcaption}[1]{\vspace{-2mm}\caption{#1}\vspace{-1mm}}
\newcommand{\vparagraph}[1]{\vspace{-2mm}\paragraph{#1}\vspace{0mm}}

\title[Visual Highlighting for Situated Brushing and Linking]%
      {Visual Highlighting for Situated Brushing and Linking}

\author[N. Doerr, B. Lee, K. Baricova, D. Schmalstieg \& M. Sedlmair]
{\parbox{\textwidth}{\centering
    Nina Doerr\orcid{0000-0003-3249-5354},
    Benjamin Lee\orcid{0000-0002-1171-4741},
    Katarina Baricova\orcid{0009-0007-6841-5199},
    Dieter Schmalstieg\orcid{0000-0003-2813-2235},    
    Michael Sedlmair\orcid{0000-0001-7048-9292}
}
        \\
{\parbox{\textwidth}{\centering
         University of Stuttgart, Germany
       }
}
}
\begin{document}

% Changes \autoref so that the referenced content is always capitalised at the start, and that it only uses Section and not subsection, etc.
\renewcommand{\figureautorefname}{Figure}
\renewcommand{\tableautorefname}{Table}
\renewcommand{\partautorefname}{Part}
\renewcommand{\appendixautorefname}{Appendix}
\renewcommand{\chapterautorefname}{Chapter}
\renewcommand{\sectionautorefname}{Section}
\renewcommand{\subsectionautorefname}{Section}
\renewcommand{\subsubsectionautorefname}{Section}

\teaser{
 \vspace{-1.2cm}
 \includegraphics[width=0.85\linewidth]{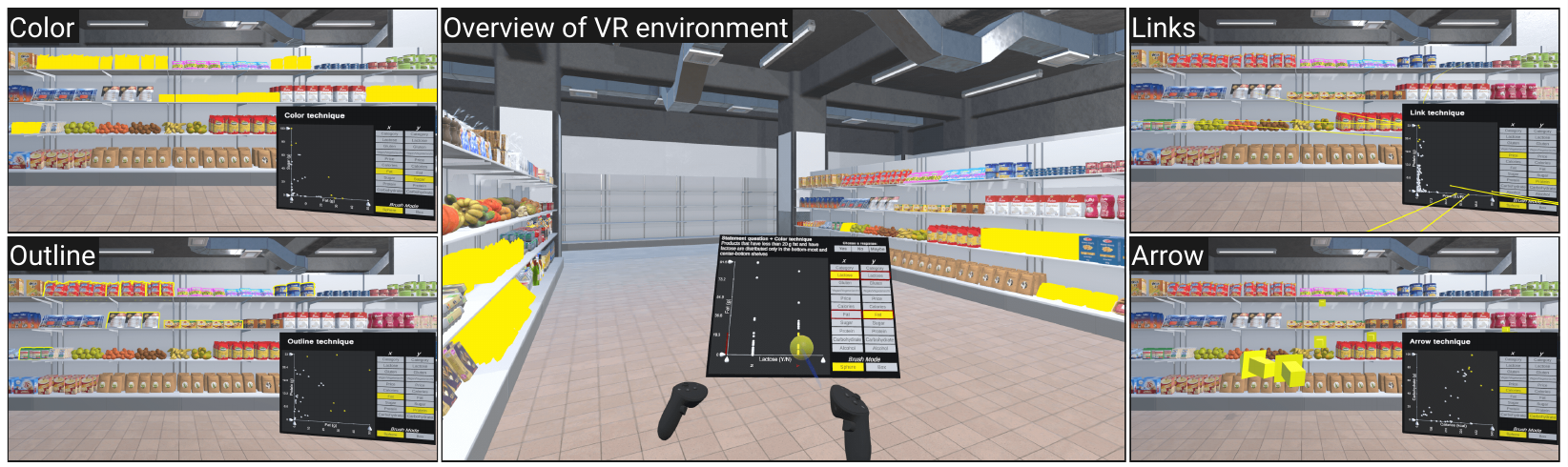}
 \centering
 \vspace{-0.15cm}
 \caption{In our user study, we present a virtual \supermarket with four highlighting techniques. Center: First-person view of the \supermarket while brushing a scatterplot, with products on the shelves highlighted. Top left: The \col technique. Bottom left: The \outline technique. Top right: The \link technique with products behind the camera selected. Bottom right: Still image of the \arrow technique.}
\label{fig:teaser}
}

\maketitle
%-------------------------------------------------------------------------
\begin{abstract}
   Brushing and linking is widely used for visual analytics in desktop environments. However, using this approach to link many data items between situated (e.g., a virtual screen with data) and embedded views (e.g., highlighted objects in the physical environment) is largely unexplored. To this end, we study the effectiveness of visual highlighting techniques in helping users identify and link physical referents to brushed data marks in a situated scatterplot. In an exploratory virtual reality user study (N=20), we evaluated four highlighting techniques under different physical layouts and tasks. We discuss the effectiveness of these techniques, as well as implications for the design of brushing and linking operations in situated analytics.
%-------------------------------------------------------------------------
%  ACM CCS 1998
%  (see https://www.acm.org/publications/computing-classification-system/1998)
% \begin{classification} % according to https://www.acm.org/publications/computing-classification-system/1998
% \CCScat{Computer Graphics}{I.3.3}{Picture/Image Generation}{Line and curve generation}
% \end{classification}
%-------------------------------------------------------------------------
%  ACM CCS 2012
%   (see https://www.acm.org/publications/class-2012)
%The tool at \url{http://dl.acm.org/ccs.cfm} can be used to generate
% CCS codes.
%Example:
\begin{CCSXML}
 <ccs2012>
    <concept>
        <concept_id>10003120.10003145.10011769</concept_id>
    </concept>
    <concept>
        <concept_id>10003120.10003121.10011748</concept_id>
        <concept_desc>Human-centered computing~Empirical studies in HCI</concept_desc>
        <concept_significance>500</concept_significance>
    </concept>
    <concept>
        <concept_id>10003120.10003145.10003147.10010923</concept_id>
        <concept_desc>Human-centered computing~Information visualization</concept_desc>
        <concept_significance>300</concept_significance>
   </concept>
</ccs2012>
\end{CCSXML}

\ccsdesc[500]{Human-centered computing~Empirical studies in visualization}
\ccsdesc[500]{Human-centered computing~Empirical studies in HCI}
\ccsdesc[300]{Human-centered computing~Information visualization}
\printccsdesc   
\end{abstract}  
%-------------------------------------------------------------------------
\vsection{Introduction} 
\label{sec:introduction}
Situated analytics (SitA) involves the use of situated visualizations \cite{whiteSiteLensSituatedVisualization2009} to improve sense-making in physical contexts \cite{elsayedSituatedAnalytics2015, thomasSituatedAnalytics2018}. It is facilitated predominantly by the use of augmented reality (AR) \cite{bressaWhatSituationSituated2022}. Willett et al.~\cite{willettEmbeddedDataRepresentations2017} distinguish visualizations directly embedded with one particular physical referent from situated visualizations that indirectly link to a referent or the entire environment. Shin et al.~\cite{shinRealitySituationSurvey2023} note that most current SitA tools use the referent as a trigger to instantiate and modify visualizations. For instance, \textit{scanner} applications augment real-world objects with additional information as they come into view. SitA is therefore reliant on the user's viewpoint and interactions with the real world to determine what information to display.

Although this ``physical world first'' approach feels natural in the context of AR, it is inherently constrained in situations with many physical referents or referents that are spread far apart \cite{leeDesignPatternsSituated2023}. Consider the classic \supermarket scenario introduced by ElSayed et al.~\cite{elsayedSituatedAnalyticsDemonstrating2016}, which features each product in the store having its own embedded visualization. While this scenario works well for products in view, it does not scale well to products \textit{outside} of one's view. Even with out-of-view labeling techniques~\cite{gruenefeldFlyingARrowPointingOutofView2018,linLabelingOutofViewObjects2023}, the customer in one-to-one configuration would still need to look at and interpret the visualization of each individual product \cite{leeDesignPatternsSituated2023}. The issue is exacerbated by the small field of view (FOV) of current state-of-the-art AR headsets.

Situated visualizations that are presented independently of individual referents can naturally consolidate the data pertaining to many referents in one view or dashboard \cite{Ma2021,sayara2023designing}. Instead of looking at multiple embedded visualizations, a single situated overview visualization can be looked at instead. However, should the task require the identification or physical manipulation of referents (e.g., when shopping), then supporting the \textit{transition} between a situated visualization of all referents to an individualized, embedded view of specific referents is required \cite{willettEmbeddedDataRepresentations2017}.

We postulate that this problem is functionally equivalent to \textit{brushing and linking}, a fundamental technique in visual analytics \cite{Roberts2007}. Selections made in the situated visualization must be visually linked to the corresponding physical referents. Linking therefore serves both analytic (making selections based on certain criteria) and perceptual purposes (finding and locating objects).

In this paper, we explore multiple techniques to visually highlight physical referents in the context of brushing and linking. We focus on the \textit{linking} between a situated visualization and its physical referent, rather than the interactions used to perform the \textit{brushing}, as it is the visual linking that fundamentally facilitates the transition between views. While existing work has investigated many ways to highlight and guide attention in 3D environments (see \autoref{sec:related-work}), to the best of our knowledge, none have explored highlighting for brushing and linking in SitA.
We selected four highlighting techniques known from either traditional brushing and linking or attention guidance in AR/VR. We conducted an exploratory user study (N=20), using a \supermarket scenario. The study was carried out in virtual reality (VR) to avoid hardware and tracking limitations of current AR devices. 
In summary, we contribute:
\begin{itemize}
    \item An open-source VR prototype demonstrating the use of brushing and linking in a situated analytics scenario
    \item An exploratory user study that seeks to compare, evaluate, and understand four common highlighting techniques
    \item Observations and lessons learned for the application of highlighting techniques in situated brushing and linking contexts
\end{itemize}

%-------------------------------------------------------------------------
\vsection{Related work} 
%%%%%%%%%%%%%%%%%%%%%%%%%%%%%%%%%%%%%%%%%%%%%%%%%%%%%%%%%%%%%%%%%%%%%%%%%%%
\label{sec:related-work}

First, we consider \textit{situated analytics} (\autoref{sec:related-work-sita}) and, second, \textit{brushing and linking} (\autoref{sec:related-work-bandl}) and its use in immersive and situated environments. As we focus on the linking component, we also discuss \textit{attention guidance in AR and VR} (\autoref{sec:related-work-attentionguidance}).

\vsubsection{Immersive and situated analytics}
\label{sec:related-work-sita}

Immersive technologies, both VR and AR, are increasingly being used for immersive analytics~\cite{marriottImmersiveAnalytics2018} of spatial data~\cite{Fonnet2019} and abstract data~\cite{Kraus2022}. Research has shown how the 3D environment can be used as an ``immersive space to think''~\cite{lisleSensemakingStrategiesImmersive2021}, populating this space with multiple views and representations~\cite{Roberts2022}. However, care must be taken to consider the pitfalls of designing for immersive displays~\cite{McIntire2014}. Designers of VR interfaces for analytics must take into account the characteristics of VR displays that differ from desktop computing, such as stereoscopic depth, angular resolution, field of view, or fatigue related to physical navigation and interaction~\cite{Laviola2017}.  

Recent work considers immersive analytics techniques with AR in physical environments, a new direction referred to as situated analytics (SitA)~\cite{elsayedSituatedAnalytics2015, elsayedSituatedAnalyticsDemonstrating2016}. SitA is characterized by the ability to perform higher-level analytical reasoning aided by the user's physical environment~\cite{shinRealitySituationSurvey2023}, particularly when there is a semantic relationship between the data and its physical referents~\cite{whiteSiteLensSituatedVisualization2009, willettEmbeddedDataRepresentations2017}. While research has investigated how situated visualizations can be designed for SitA~\cite{leeDesignPatternsSituated2023,shinRealitySituationSurvey2023}, considering human perception to leverage efficient visual processing---particularly in situated contexts---is a grand challenge~\cite{Ens2021}.

Designers of SitA systems (and AR interfaces in general) must handle the additional constraint that visual representations cannot always be positioned on a ``clean slate'' background. Kruijff et al.~\cite{Kruijff2010} list several challenges related to visual perception in AR which are caused by the environment, image capture, display, and visual representation. Satkowski et al.~\cite{Satkowski2021} study how the real-world background affects perception in AR, and Assor et al.~\cite{assorHandlingNonVisibleReferents2024} propose a design space of how to handle non-visible physical referents. Similarly, AR must always consider the interplay of visual representation with the existing and immutable appearance of the real world. Therefore, any visualization pipeline targeting AR needs to deal with the fusion of real and virtual visual attributes~\cite{Zollmann2021}. Addressing these challenges requires a wide range of visual encoding methods~\cite[chapter 7]{Schmalstieg2016}.

\vsubsection{Brushing and linking}
\label{sec:related-work-bandl}
One of the core topics of this paper is brushing and linking~\cite{Roberts2007}, which is the process of locating the corresponding items of interest in multiple views. Fundamentally, it comprises two sequential stages, wherein selections made in one visualization (brushing) are automatically shown in another visualization (linking)~\cite{keimInformationVisualizationVisual2002}. Much research has focused on the brushing component, investigating interaction techniques to support the effective selection of data points in desktop computing \cite{koytekMyBrushBrushingLinking2018}.

In immersive analytics, brushing and linking have seen (arguably limited) use, be it to compare multiple immersive coordinated views~\cite{ahnSupportingHealthyGrocery2015,liuDesignEvaluationInteractive2020} or to support collaborative awareness~\cite{leeSharedSurfacesSpaces2021, saffoTheirEyesTheir2023,serajiXVCollabImmersiveAnalytics2022}. Most methods use simple 3D selection techniques (e.g., raycasting) or pure 2D selection techniques (e.g., touchscreen inputs). While we do not propose new brushing interaction techniques, we are the first (to the best of our knowledge) to explore brushing and linking in a SitA context---particularly when using physical referents as a coordinated view.

\vsubsection{Attention guidance in augmented and virtual reality}
\label{sec:related-work-attentionguidance}

In SitA, we assume that the linking component targets views and referents in the real world. Hence, it can be described as a visual search task for target objects in an AR display. Note that this work considers attention guidance as a means to support visual search and identification, and not as instructional cues for movement training \cite{yuDesignSpaceVisual2024} or assembly tasks \cite{pietschmannQuantifyingImpactXR2023}. Wolfe et al.~\cite{Wolfe2011} characterize visual search as the task of visually identifying a target among other, irrelevant objects (distractors). If there are too many distractors, a naive visual search becomes inefficient \cite{Vertegaal2002}. We must therefore ensure that the visual search is tolerant of clutter and occlusion, since we cannot change the real world. There should be sufficient contrast and legibility between real and virtual objects. We have to be able to direct a user's attention to the relevant objects, in particular, if these are out of view.

According to Cockburn et al.~\cite{Cockburn2009}, conventional InfoVis methods to guide visual attention and search, such as overview + detail or focus + context, occupy a lot of screen space as they enlarge objects or introduce replicas of different sizes. As a major alternative, they mention cue-based approaches, which modify object rendering styles. Lin et al.~\cite{linLabelingOutofViewObjects2023} note that such cues are preferred in AR, since resizing or duplicating occupies too much of the natural view space. These cue-based techniques for highlighting focal objects can select from a variety of rendering attributes. A baseline is formed by highlights, which emphasize a target by modifying its color throughout the target area or as an outline \cite{koytekMyBrushBrushingLinking2018}. 

In immersive displays, several more sophisticated techniques have been proposed to ensure that a given contrast is achieved~\cite{Grogorick2018, Kalkofen2013}. Contrast modulation, even below a detectable threshold \cite{Veas2011}, can pre-attentively guide the user. Other attributes that can be modified to highlight or attract attention include spatial frequencies (i.e., focus or blur)~\cite{Kosara2001}, stereoscopy~\cite{Krekhov2020}, or temporal frequencies~\cite{Lange2020}.

Cockburn et al.~\cite{Cockburn2009} further mention proxies as cues. Arrow-shaped proxies are commonly used in AR \cite{gruenefeldFlyingARrowPointingOutofView2018, Tonnis2006}, most likely because they are simple to synthesize and easy to understand. Naturally, proxies are a popular means to guide toward out-of-view targets. In addition to arrows, various popular shapes include halos \cite{Baudisch2003}, funnels \cite{Biocca2006}, compass-like~\cite{Suomela2000} or radar-like widgets~\cite{Jo2011}, and lines or trajectories~\cite{prouzeauVisualLinkRouting2019}. 
Previous work also compares the effect of various visual cues in VR and AR. Several authors~\cite{Bork2018, Kishishita2014, Trepkowski2019, WielandJ2022} investigate the performance differences of visual search with respect to the field of view of the display. Other studies compare multiple attention guidance (i.e., highlighting) techniques with respect to the effectiveness and preservation of immersion~\cite{Lange2020}, the effectiveness of multimodal cues~\cite{Marquardt2020}, or the influence of stationary and moving distractors~\cite{Doerr2023}. Moreover, our selection of techniques was influenced by the work of Whitlock et al.~\cite{Whitlock2020} on the perception of fundamental graphical attributes in immersive analytics. 
However, none of these works investigate the use of AR or VR attention guidance in the context of brushing and linking. Therefore, our work seeks to bridge the gap between these two topics necessary to enable brushing and linking in SitA.

%-------------------------------------------------------------------------
\vsection{Experimental prototype for situated brushing and linking}
%%%%%%%%%%%%%%%%%%%%%%%%%%%%%%%%%%%%%%%%%%%%%%%%%%%%%%%%%%%%%%%%%%%%%%%%%%%
\label{sec:prototype}

This research aims to investigate how visual highlighting could support brushing and linking in SitA. As mentioned in \autoref{sec:related-work}, brushing and linking, as well as visual highlighting (i.e., attention guidance) are not novel ideas when considered individually, but it is the intersection in SitA that requires further study. 
Fortunately, existing techniques can be adapted to SitA. To ensure a more grounded and systematic approach, we draw upon popular techniques already used in brushing and linking in visual analytics, and attention guidance in AR and VR. This approach has dual benefits. First, we can immediately see if prior research directly translates into SitA. Second, we can establish a baseline understanding of brushing and linking in SitA without the confounding effects of potentially complex highlighting techniques.
In this section, we describe a VR prototype that we developed to investigate situated brushing and linking, which is open source on GitHub~\cite{VisHigh}.

\vsubsection{Scenario}
SitA is relevant in many situations involving data and its physical context \cite{bressaWhatSituationSituated2022}. Brushing and linking becomes useful when considering the actual appearance or location of physical referents. We therefore chose a \supermarket scenario to situate our experimental prototype and the subsequent user study (\autoref{fig:teaser}). A \supermarket is a popular scenario in the SitA literature~\cite{ahnSupportingHealthyGrocery2015, elsayedSituatedAnalytics2015, elsayedHORUSEYESee2016}. Its defining feature is an abundance of grocery products (referents) laid out on shelves, which vary in layout, making the scenario well-suited for brushing and linking. We also considered similar environments, such as a library with books or a building fa\c{c}ade with windows as referents~\cite{Tatzgern2016}. We built the \supermarket using VR as a kind of information-rich virtual environment~\cite{Bowman2003, Polys2006}, serving as a simulated test environment that can be carefully controlled and replicated~\cite{Ragan2009}. While using AR would be more ecologically valid, the use of VR avoids the technical challenges of tracking each product~\cite{Sousa2023} or logistical challenges of renting or building a store. Previous work also suggested that findings in AR can be replicated in VR \cite{GrandiJ2021Workshop,21LeeC2009Replication,22LeeC2010RoleofLatency,23LeeC2013Replication}, which further motivates and validates our use of a simulated \supermarket in VR.

\begin{figure*}
    \centering
    \includegraphics[width=\textwidth]{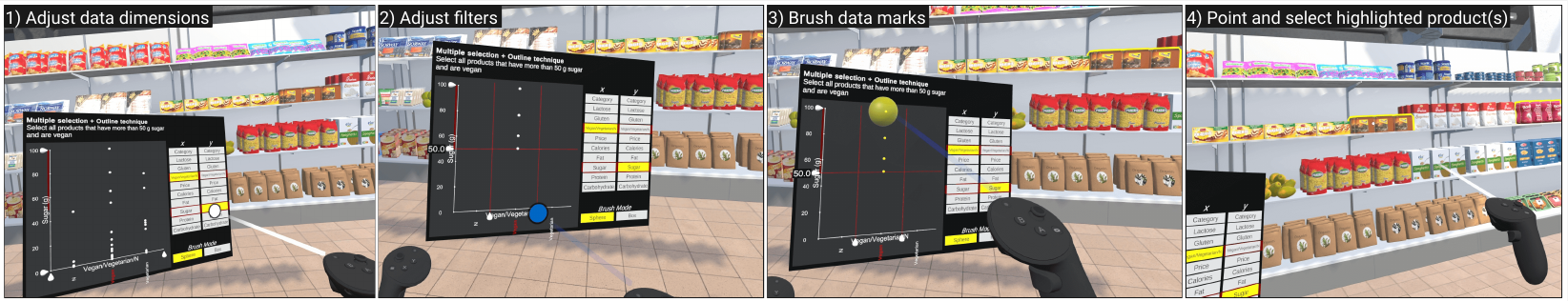}
    \vspace{-0.425cm}
    \vcaption{Sequential frames of brushing on a situated scatterplot and identifying linked products (referents) during our user study, with the virtual tablet being held at head height for illustrative purposes.
    Frame 1: The user adjusts the data dimensions as desired. Frame 2: The user adjusts the filters to show only the data marks within the desired ranges. Frame 3: The user selects the data marks. Frame 4: The user finds the highlighted product(s) and confirms identification by pointing and selecting with controller.}
    \label{fig:brushing-steps}
    \vspace{-0.25cm}
\end{figure*}

\vsubsection{Highlighting techniques}
\label{sec:highlighting-techniques}
We implemented four different highlighting techniques in our prototype. These techniques are used in conjunction with the accompanying \textit{brushing} component, which is described in \autoref{sec:brushing-interactions}. Example images of all four techniques can be seen in \autoref{fig:teaser}. Note that all techniques use a solid yellow color where relevant in order to ensure a consistent appearance across conditions.

\textbf{Color} is likely the de facto highlighting technique used in regular brushing and linking~\cite{koytekMyBrushBrushingLinking2018,Roberts2007}. It mainly benefits from the pre-attentive ``popout'' effect, particularly when the new color highly contrasts with the background. Hence, we incorporate it into our prototype, changing the color of the product to a solid yellow when selected. While we had initially tried to use a semi-transparent yellow tint on the products instead, pilot tests showed that this mode was not able to sufficiently ``pop-out'' specific products: A red cereal box, when tinted yellow, simply looks like an orange cereal box, and not a red one that is highlighted.

\textbf{Outline} is another common technique found in brushing and linking \cite{geymayerShowMeInvisible2014,gratzlDominoExtractingComparing2014}. Moreover, it is also a popular technique used in visual highlighting to show that an object is ``selected''~\cite{Dillman2018, sidenmarkOutlinePursuitsGazeassisted2020}. Although \col and \outline have similar characteristics in how they highlight targets, the actual appearance of a product is visible for \outline but not for \col. Sometimes a more subtle, non-obtrusive highlighting can be beneficial (e.g., for safety). We show outlines on the borders of the selected grocery products in camera space. We chose a width for the outline which, after pilot testing, we felt was sufficiently wide to see the outline, but thin enough not to obstruct neighboring products.  

\textbf{Links} are trajectories used to draw connections between entities~\cite{collinsVisLinkRevealingRelationships2007}. In the context of brushing and linking, a link indicates that two connected data marks are the same underlying data record \cite{koytekMyBrushBrushingLinking2018}. The same premise holds in SitA, where a data mark on the situated visualization (i.e., the virtual tablet in \autoref{fig:teaser}) is explicitly connected to its associated referent. Particularly for the comparison and analysis across view, direct connections can facilitate tasks. To this end, we leverage a prior Unity implementation by Prouzeau et al.~\cite{prouzeauVisualLinkRouting2019} which draws visual links between pairs of objects in a 3D scene. We adjusted some of their script's parameters to make the links more responsive to movement, because we allow the situated visualization to be moved by the user.

\textbf{Arrows} are ubiquitous symbols used to direct attention. They are popular for attention guidance in AR, particularly towards objects or points of interest that are out of view \cite{kasaharaJackInIntegratingFirstperson2014, Tonnis2006, WielandJ2022,YuD2020}. Thus, we consider how similar arrows could be used to highlight objects in SitA. For our study, we extended \textit{FlyingARrow}~\cite{gruenefeldFlyingARrowPointingOutofView2018}, which repeatedly instantiates 3D arrows in front of the user that then fly toward a target object. Static arrows like Wieland et al.~\cite{WielandJ2022} or Yu et al.~\cite{YuD2020} spawn in the viewport, causing clutter when there are multiple targets and hence might be less beneficial in real-life scenarios than dynamic arrows~\cite{gruenefeldFlyingARrowPointingOutofView2018}. We also suspect that adding animations or motion makes targets more salient. Since Gruenefeld et al.~\cite{gruenefeldFlyingARrowPointingOutofView2018} claim a low workload, we suspected the dynamic arrow to be a reasonable choice for highlighting in SitA. Our adapted version of FlyingARrow is able to present multiple arrows concurrently and dynamically change the arrows to support on-the-fly brushing.

\textbf{Alternatives} beyond these four fundamental highlighting techniques were considered but excluded after preliminary testing. \textit{Size} was an obvious candidate~\cite{gutwinPeripheralPopoutInfluence2017}, but it became apparent that expanding objects in the real world is impractical because they take up too much of the physical space~\cite{linLabelingOutofViewObjects2023}. \textit{Semantic depth of field}~\cite{Kosara2001} was another candidate, as blur can intentionally obscure unimportant parts of the scene. However, we soon realized that blur is not only visually distracting, but also potentially dangerous when certain parts of the real world are visually deteriorated.

\vsubsection{Brushing interactions}
\label{sec:brushing-interactions}
As the premise of our work involves both \textit{brushing} and \textit{linking}, our prototype needs faculties to enable brushing in a situated visualization. A virtual tablet attached to the non-dominant-hand controller (\autoref{fig:teaser}, center) presents a 2D scatterplot, generated with DXR \cite{sicatDXRToolkitBuilding2019}. The $x$ and $y$ data dimensions of the scatterplot are determined by buttons on the right side of the tablet.
The tablet's size (40~cm by 31~cm) ensures that its contents are easily readable in VR, while still small enough to not be visually obtrusive.

We opt for a simple brushing approach, reminiscent of the brushing mechanism used in FIESTA~\cite{leeSharedSurfacesSpaces2021}. Brushing is performed by pointing the dominant-hand controller at the scatterplot and pressing either the index button to add to the selection or the grip button to remove it from the selection. We provide two modes for brushing: spherical and rectangular. The mode can be changed using the corresponding buttons on the tablet. Any brushed data marks on the scatterplot turn a solid yellow. To make selections even easier, we added simple range filters along the two data axes. These filters are represented as handles which can be dragged using raycast interactions, similar to ImAxes~\cite{cordeilImAxesImmersiveAxes2017}. Filtered data marks cannot be brushed, ensuring that no unwanted points are included. Step-by-step images of this situated brushing and linking process are shown in \autoref{fig:brushing-steps}. The tablet could be held comfortably at hip height without obscuring any referents, as was the case for everyone during our pilot testing and subsequent user study. As our intention is to evaluate the effectiveness of the four techniques to highlight referents, we focus on the unidirectional brushing from the situated visualization to the referents.

%-------------------------------------------------------------------------
\vsection{User study}
%%%%%%%%%%%%%%%%%%%%%%%%%%%%%%%%%%%%%%%%%%%%%%%%%%%%%%%%%%%%%%%%%%%%%%%%%%%
\label{sec:methods}
We conducted a user study with our VR prototype to determine how the highlighting techniques presented in \autoref{sec:highlighting-techniques} perform under different referent layouts as well as situated brushing and linking tasks. The overall goal of the study was to determine: (1) how well existing techniques from brushing and linking and attention guidance hold up in SitA; (2) how well each highlighting technique supports our user experience categories (\autoref{sec:measures}); and (3) how brushing and linking in SitA can potentially be improved.

\vsubsection{Study conditions}
Our user study involved two independent variables:

\textbf{Highlighting technique.} \col, \outline, \link, and \arrow, as described in \autoref{sec:highlighting-techniques}.

\begin{figure}
    \centering
    \includegraphics[width=\linewidth]{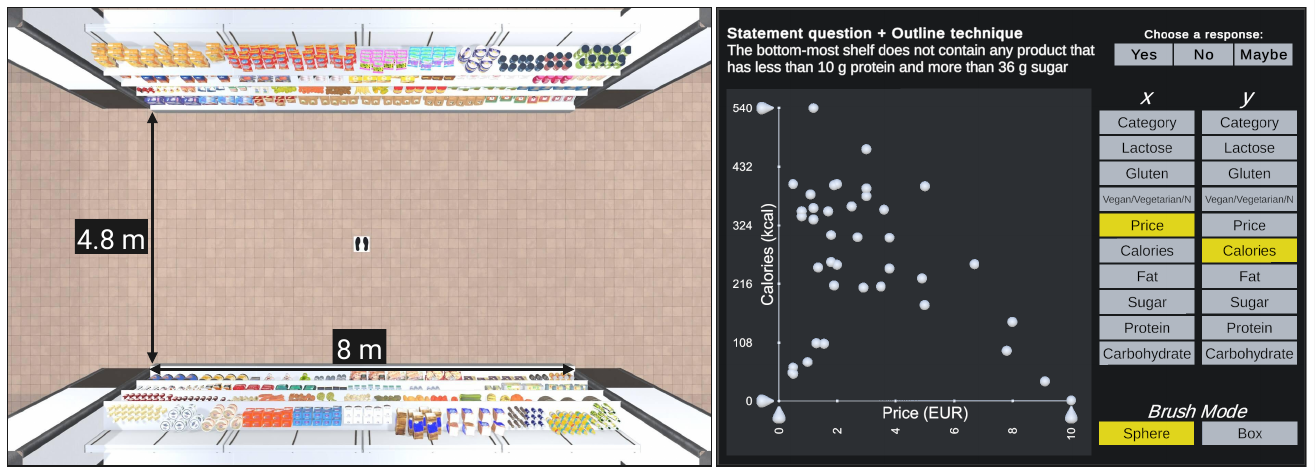}
    \vspace{-0.425cm}
    \vcaption{Left: A top-down view of the virtual environment in which the study takes place. Right: A close-up view of the panel and the user interface prompting a user response.}
    \label{fig:study-overview}
    \vspace{-0.35cm}
\end{figure}

\textbf{Shelf layout.} \textit{Inside-FOV} or \textit{Outside-FOV}. For \textit{Inside-FOV}, there are only four shelves in front of the user. \textit{Outside-FOV} extends the layout by adding four shelves behind the user, with front and back rows spaced 4.8~m apart. This extended layout emulates situations where the highlighted referents may be positioned outside of the user's FOV. Each shelf is 2~m wide by 2~m tall, forming an overall aisle with a width of 8~m in total (\autoref{fig:study-overview} left). Neither layout forces the user to move to resolve occlusion. This choice eliminates the need to support any special VR locomotion techniques, which would probably have confounded our results.

\vsubsection{Tasks and data}
Our study involved three main task types: single-selection, multi-selection, and statement response. All tasks required participants to first make a selection on the scatterplot and then look for one or more highlighted products on the \supermarket shelves. All tasks are bivariate, with the scatterplot selection being conditioned in both dimensions. The tasks are intentionally kept low-level to allow for generalizability across various SitA applications involving multiple physical referents, such as browsing cars based on mileage and price, or planning seating arrangements at a wedding based on guests' age and relationship to the bride and groom.

\textbf{Single-selection tasks.} 
Identify, select, and locate a single product (referent) that matches a bivariate criterion, e.g., \textit{``Select the product that has the highest fat and no sugar.''} This task helps determine a technique's effectiveness in locating a single referent.

\textbf{Multi-selection tasks.} 
Identify, select, and locate all products that match a bivariate criterion, e.g., \textit{``Select all products that have more than 22.5 g protein and less than 9.5 g fat.''} This task helps determine a technique's effectiveness in locating multiple referents, as it requires highlighting to be applied to multiple products simultaneously. All questions have five products that match the given criterion to ensure consistency across trials.

\textbf{Statement response tasks.}
Respond with ``Yes'', ``No'', or ``Maybe'' to a given statement that describes the dataset and \supermarket layout, e.g., \textit{``Vegan products that contain more than 51 g of carbohydrate are distributed only in the center-left and center-right shelves.''} This task requires participants to observe a collection of referents together and make a judgment on their spatial arrangement. The task is also reminiscent of standard brushing and linking, wherein all brushed points should be visible and considered when judging correlations and dependencies in the linked view \cite{keimInformationVisualizationVisual2002}.

We used a curated dataset that contains 11 variables (7 quantitative, 4 categorical) for 98 products, which we placed on the \supermarket shelves. This data was manually scraped from \supermarket websites, and all values were checked for plausibility. Minor adjustments were made to these values to accommodate the above tasks. The \textit{Inside-FOV} condition only uses a reduced dataset of 39 products fitting in the front-facing set of shelves.

\vsubsection{Experimental setup}
We used the experimental prototype described in \autoref{sec:prototype}. All instructions and buttons used to answer the statement response tasks (``Yes'', ``No'', ``Maybe'') are shown in the upper area of the tablet (see \autoref{fig:study-overview} right). Additionally, functionality was added for the single- and multi-selection tasks to indicate that they have ``identified'' the required products. 
%
% not so interesting
As we are not interested in the usability of brushing scatterplots in VR, we decided to streamline the process after pilot testing revealed it to be too cumbersome. Red visual cues were added to indicate the two required data dimensions for each task and to indicate the required value ranges along the axes. The filter handles were modified to ``snap'' to these range thresholds. This simplification does not confound our primary focus on the ``linking''. \autoref{fig:brushing-steps} shows the steps involved in solving a multi-selection task, including indicating identified referents. In terms of movement, we neither forced participants to move nor restricted their movement. When asked for clarification, we told them that all products could be seen just by turning.

We conducted our study using a desktop PC equipped with an Intel Core i9-9980XE CPU, RTX 2080 Ti with 11~GB VRAM, and 128~GB of RAM. We used an Oculus Quest Pro headset with two hand controllers. The headset was connected via a 5~m Quest \link USB cable to the PC running our application developed in Unity version 2022.2.20f1. All questionnaires were administered on the same PC using LimeSurvey. The study was carried out in an open space of 360~cm by 220~cm that was free of obstructions.

\subsection{Measures} \label{sec:measures}
\hspace{\parindent}\textbf{Task Performance.} 
We measured task performance using completion time and errors. For each task, we recorded the overall completion time, the time of the last interaction with the tablet (i.e., button presses, filter changes, and brushing), and the time when the first product was selected (\autoref{fig:brushing-steps}, frame 4). The overall completion time is the interval between starting the task and selecting the final product or responding to the given argument via the respective button. Incorrectly selected products are counted as errors.

\textbf{User Experience.} 
We measured user experience using two questionnaires. A questionnaire was completed after each shelf layout for each technique, and another, final questionnaire, after all conditions. The in-between questionnaire included \textit{NASA-TLX}~\cite{nasatlx2006}, free-text forms for comments on user strategies and experiences, as well as additional feedback. The physical environment might not be as accessible as the desktop visualizations and might introduce additional factors that influence certain highlighting techniques. Hence, we came up with the following categories that such techniques should support through internal discussions, our own experiences and findings with SitA, and reported findings in the literature. The final questionnaire included these categories using 9-point Likert scales (1=worst, 9=best, unless stated otherwise).

\textbf{General experience.} A subjective rating of how the highlighting technique performed in general.

\textbf{Out-of-view.} A highlighting technique ideally accounts for physical referents being out of view or occluded, since identifying can be equally challenging for virtual and physical objects.

\textbf{Visual clutter.} A highlighting technique ideally supports marking multiple referents without clutter, since scalability is often desired for both infovis~\cite{richer2022scalability} and SitA~\cite{buttnerAugmentedRealityTraining2020}.
 
\textbf{Obstruction.} A highlighting technique ideally minimizes occluding its surrounding and obstructing the task, as distracting or reducing the user's real world awareness can be dangerous (1=not at all, 9=very).

\textbf{Subtlety.} A highlighting technique ideally balances between saliency and subtlety, as some SitA contexts (e.g., for safety) require a more subtle highlighting technique (1=very subtle, 9=very obtrusive).

\textbf{Explicit connections.} A highlighting technique ideally supports direct connections of data marks and referents, since making assumptions across views is common in visual analytics~\cite{Roberts2007} and potentially in SitA scenarios as well.

\textbf{Recognition.} A highlighting technique ideally balances between identifying and obscuring the target, as some real-world tasks (like grocery shopping) might require to be aware of the actual appearance while others do not (1=not at all, 9=very).

\textbf{Enjoyment.} A highlighting technique balances between usability and enjoyment resulting from an aesthetically pleasing design.

The relevance of the categories varies between use cases. However, we collected all categories throughout the experiments to obtain a full characterization of possible situated brushing and linking contexts. The final questionnaire also asked participants to describe their strategies for different techniques and shelf layout conditions, and provide general feedback.

\vsubsection{Study design and procedure}
We used a within-subjects study design to evaluate our two independent variables: highlighting technique (\col, \outline, \link, \arrow); and shelf layout (Inside-FOV, Outside-FOV). Each condition consisted of three tasks, one per each task type (single-selection, multi-selection, statement response). Each question was randomly selected from a set of eight predefined questions per type of task. Thus, each participant had to perform four highlighting techniques $\times$ two shelf layouts $\times$ three tasks = 24 trials. We counterbalanced the order of the highlighting technique using Latin squares. However, we used a fixed order for both the shelf layout (Inside-FOV $\rightarrow$ Outside-FOV) and the task type (single $\rightarrow$ multi $\rightarrow$ statement), because these two aspects have a naturally increasing level of difficulty. Participants were compensated with €12.

\textbf{Introduction (5 minutes).} The participant was welcomed and introduced to the purpose of the study. After signing a consent and privacy form, they were asked for their handiness, asked to fill in a demographic questionnaire, and introduced to the VR headset and controllers. They were informed that they could take breaks or withdraw at any point in the study without consequences.

\textbf{Training (10-15 minutes).} The experimenter first explained the VR prototype to the participant, describing the \supermarket and tablet, the available interactions and the tasks they would be performing. PowerPoint slides with pictures were also shown during the briefing. The participant then put on the VR headset and practiced brushing and linking interactions using a simplified layout of 11 products and the aforementioned \textit{size} highlighting technique. In this phase, no task was assigned, and the participant could freely try the interactions without time pressure.

\textbf{Main study (40-50 minutes).} The participant was then given each of the conditions and tasks in the counterbalanced order described above. Whenever a new highlighting technique was introduced, the participant was given a short tutorial using another simplified layout of four products to let them see what the highlighting technique looked like. Between each trial, the participant was told to return to the center of the room and face the direction indicated by the image on the floor (see \autoref{fig:study-overview} left). After completing each set of three tasks (single-selection, multi-selection, statement response), the participant removed the VR headset to answer the in-between questionnaire. After a short break (if needed), they put the headset back on to proceed with the study.

\textbf{Post study (5-10 minutes).} After all trials were completed, the participant answered a final questionnaire. Then they were given compensation as a reward for their participation.

\vsubsection{Guiding questions}
For the analysis of our exploratory study, we devised four guiding questions, along with some of our prior expectations:
\begin{itemize}
    \item \textbf{GQ1: Which techniques performs the best in terms of task performance, and which the worst?} We expect that \link will have the fastest completion time due to its guidance effect, even to out-of-view targets, as well as its direct connection between data marks and its referents. We also expect that \arrow leads to the slowest completion time for the multi-selection task due to its limited scalability. We have no expectations about error rates. 
    \item \textbf{GQ2: Which technique is subjectively preferred or disliked?} \link fits many required features of our tasks and requirements (e.g., out-of-view highlighting or direct connection of data marks and their referents). We assume that it is perceived as the easiest and fastest technique and expect it to be preferred.
    \item \textbf{GQ3: Which technique is the least preferred if many targets are not in view?} We expect \arrow to be perceived worst, as it is not scalable and, therefore, clutters an environment containing many targets. For targets outside the view, we expect \col and \outline to be the least preferred, as they do not have guidance and no direct connections between data marks and referents.
    \item \textbf{GQ4: Which technique is perceived as the most subtle and least obstructive?} We expect \outline to be deemed the most subtle and least obstructive, as it obscures only the boundary of the targets and does not cause clutter even with many targets.
\end{itemize}

\begin{figure}[t]
    \centering
    \begin{subfigure}{\linewidth}
        \centering
        \includegraphics[width=\textwidth]{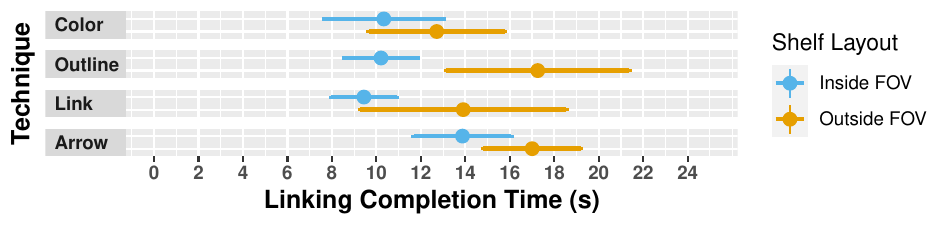}
        \vspace{-0.6cm}
        \caption{Overall linking completion time.}
        \label{fig:linkingCT}
    \end{subfigure}
    \hfill
    \begin{subfigure}{\linewidth}
        \centering
        \includegraphics[width=\textwidth]{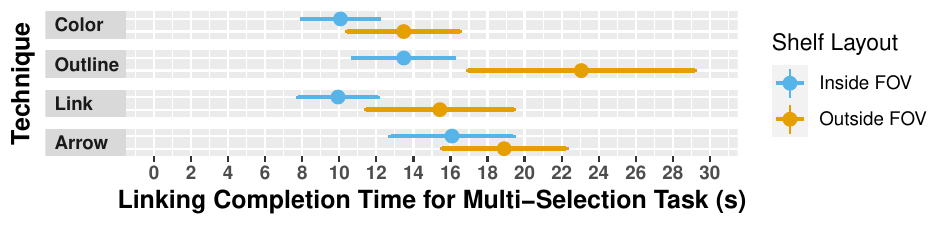}
        \vspace{-0.6cm}
        \caption{Linking completion time for the multi-selection task.}
        \label{fig:MSlinkingCT}
    \end{subfigure}
    \vspace{-0.4cm}
    \caption{Completion time measures using mean and 95\% CIs.}
    \vspace{-0.45cm}
    \label{fig:taskperformance}
\end{figure}

%------------------------------------------------------------------------
\vsection{Results}
%%%%%%%%%%%%%%%%%%%%%%%%%%%%%%%%%%%%%%%%%%%%%%%%%%%%%%%%%%%%%%%%%%%%%%%%%%%
\label{sec:results}
%demographics
We recruited 20 participants (8 female and 12 male) via university mailing lists and social media, with different age groups between 18 and 55 years (mode={26--30}). 19 participants had correct or corrected-to-normal vision, while one participant had astigmatism. We had 16 participants who were right-handed, and 4 participants who were left-handed. Most of our participants had some visualization experience (16/20 between 2 and 4 on a 5-point Likert scale), and only a few participants had little (4/20 between 0 and 1). More than half of the participants had some VR experience (12/20 between 2 and 4 on a 5-point Likert Scale), and the remaining participants had little (8/20 between 0 and 1).

%quantitative
Following the recommendations of Cumming~\cite{thenewstatistics} and the American Psychological Association~\cite{apapublicationmanual}, we visually analyze the data based on their mean and 95\% confidence intervals using forest plots. In general, the evaluation and subsequent discussion focuses more on the differences between the highlighting techniques than on the shelf layout conditions.

\begin{figure}[t]
    \centering
    \begin{subfigure}{\linewidth}
        \centering
        \includegraphics[width=\textwidth]{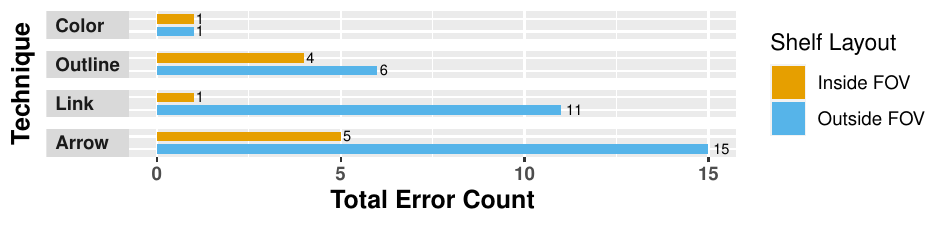}
        \vspace{-0.6cm}
        \caption{Total error counts across all participants.}
        \label{fig:errors}
    \end{subfigure}
    \begin{subfigure}{\linewidth}
        \centering
        \includegraphics[width=\textwidth]{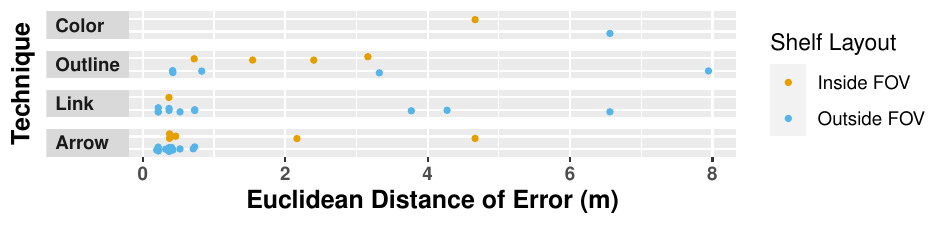}
        \vspace{-0.6cm}
        \caption{Distances between each incorrect product and its closest correct product.}
        \label{fig:errors-euclidean}
    \end{subfigure}
    \vspace{-0.6cm}
    \caption{Error measures for the single- and multi-selection task.}
    \vspace{-0.45cm}
    \label{fig:taskerrors}
\end{figure}

\vsubsection{Task performance}
Completion time and errors are shown in \autoref{fig:taskperformance} and \autoref{fig:taskerrors}.

\vparagraph{Completion Time}
Since our analysis focuses on the linking part of the task, we calculated \textit{linking completion time} defined as the overall completion time minus the time of the last interaction on the tablet. During the study, we noticed that some participants did not brush all data marks at once, but rather brushed one at a time when using certain techniques. For these cases, we manually calculated the brushing time and linking time from the recorded video.

%Overall Linking CT
For the \textbf{overall linking completion time}, which includes all tasks, we received 24 values per participant (4~techniques $\times$ 2~shelf~layouts $\times$ 3~tasks). To aggregate these values for each technique and shelf layout per participant, we first averaged the values before calculating the overall mean and the respective confidence interval. The results of the overall linking completion time can be seen in \autoref{fig:linkingCT}. The {\color{cyan}Inside-FOV} linking CT is generally faster than the {\color{orange}Outside-FOV} for all techniques. \col and \link perform similarly; \link is slightly better for the Inside-FOV, but worse for the Outside-FOV. While \outline resulted in similar results for the Inside-FOV, the results indicate a substantial performance drop for Outside-FOV targets. \outline is the only technique with a large gap between the two shelf layout conditions. \arrow is the worst for the Inside-FOV, and, second worst for Outside-FOV. 

%Multi-selection Linking CT
We also evaluated the \textbf{linking completion time for the multi-selection task}. We were specifically interested in this task, as multi-selection constitutes the "core discipline" of brushing and linking.
The results are shown in \autoref{fig:MSlinkingCT}. As before, the Inside-FOV has lower values than the Outside-FOV. Overall, \col performs best, closely followed by \link, as its Outside-FOV completion time is slightly higher. Again, \outline has a large gap between the shelf layout conditions, with the worst results for the Outside-FOV. \arrow is the worst for the Inside-FOV. By and large, the results for the multi-selection task are very similar to the overall results above, with a slightly stronger articulation of the main characteristics.

\vparagraph{Errors}
We also measured \textbf{errors} as incorrectly selected products in single- or multi-selection tasks and removed duplicate errors caused by selecting the same incorrect product repeatedly. \autoref{fig:errors} shows the total count of errors among all participants. \col had the lowest number of errors and \arrow the most, with Outside-FOV being more prone to errors, especially for \link and \arrow. In addition, \autoref{fig:errors-euclidean} shows for each error how far the correct product was in meters. From this result, we see that most of the errors made using \col and \outline were more than 1 m away, mainly due to incorrect brushing on the scatterplot. This was confirmed by reviewing the respective video footage. In contrast, for \link and \arrow, many errors were less than 0.5 m away, suggesting that participants actually mistook the product they needed to select---with some also due to incorrect brushing. Overall, \link and \arrow were more prone to errors when identifying a highlighted product.

\vsubsection{User experience}
We now report questionnaire results of the category ratings and NASA-TLX.

\begin{figure}
    \centering
    \includegraphics[width=\linewidth]{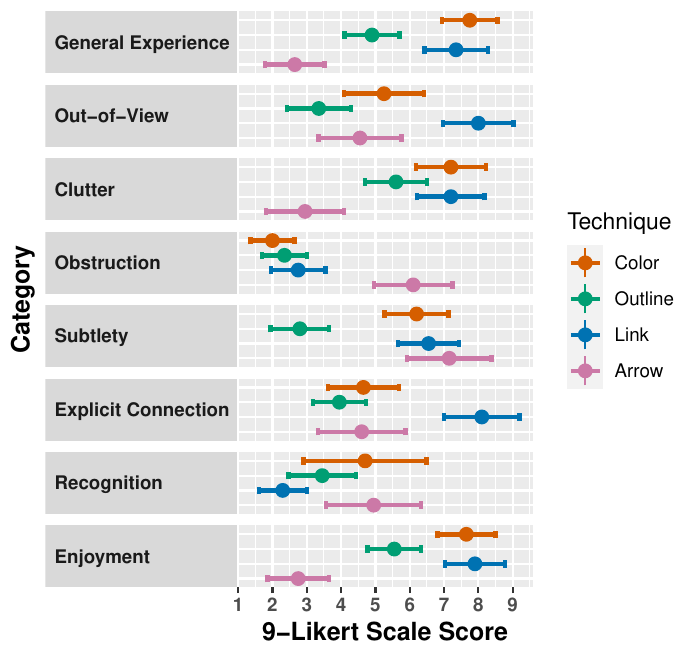}
    \vspace{-8mm}
    \caption{Mean and 95\% CIs of the category ratings.}
    \vspace{-5mm}
    \label{fig:category-rating}
\end{figure}

\begin{figure*}
    \centering
    \includegraphics[width=\textwidth]{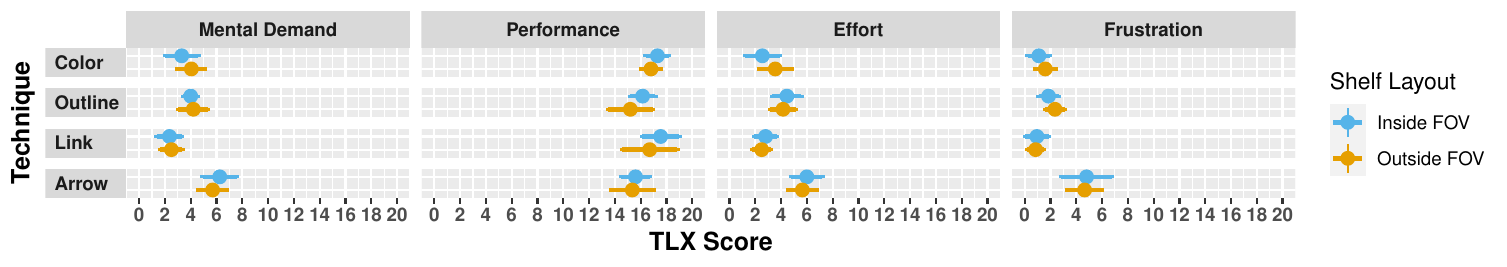}
    \vspace{-8mm}
    \caption{Mean and 95\% CIs of the NASA-TLX subscales mental demand, performance, effort and frustration.}
    \vspace{-0.45cm}
    \label{fig:important-ntlx}
\end{figure*}

\vparagraph{Category Rating}
The results of the 9-point Likert scale ratings are in \autoref{fig:category-rating}. High values are better for all categories, except for obstruction, subtlety, and recognition, where low values are better.

\begin{itemize}
    \item \textbf{General experience} {\color{orange}\col} and {\color[rgb]{0.0, 0.4, 0.65}\link} performed similarly, followed (with large gaps) by {\color[rgb]{0.16, 0.67, 0.53}\outline} and, lastly, {\color[rgb]{0.75, 0.58, 0.89}\arrow}. While \outline received a mediocre rating, \arrow was rated very low.
    \item \textbf{Out-of-view} \link was ranked the best. With a considerable gap; \col, \arrow, and \outline followed.
    \item \textbf{Visual clutter} \col and \link resulted in the best scores, followed by \outline. With a considerable gap, \arrow was the worst and considered to be the most visually cluttered. 
    \item \textbf{Obstruction} (low better): \col, \outline, and \link all performed very well, while \arrow was worst by a large margin.
    \item \textbf{Subtlety}  (low better): \outline was perceived by far as the most subtle technique. The other three were not considered subtle.
    \item \textbf{Explicit connections} \link wins with the highest score. 
    \item \textbf{Recognition} (low better): \link was judged the best for recognizing the product, followed by \outline. \col and \arrow resulted in mediocre scores. 
    \item \textbf{Enjoyment} Enjoyment results are similar to \textbf{General experience}, with the highest enjoyment for \link and \col, followed by \outline and, far behind, \arrow. 
\end{itemize}
\col and \link performed the best for all ratings. \outline was the sole favorite in \textbf{subtlety}, and \link in \textbf{explicit connection}. \arrow was the worst in six categories, and second to last in the other two.

\vparagraph{NASA-TLX}
We briefly report on four subscales of the NASA-TLX: \textbf{mental demand}, \textbf{performance}, \textbf{effort}, and \textbf{frustration} (\autoref{fig:important-ntlx}). Physical and temporal demand are not reported here as the task is not physically intensive nor does it have time constraints, though their scores were similar to mental demand. Please see the supplemental material for the remaining NASA-TLX data.

For \textbf{performance}, Inside-FOV scores are slightly higher than Outside-FOV scores but are still comparable. \col and \link have similar high scores, followed by \outline and \arrow. Overall, there are no large differences between the techniques. The remaining results for \textbf{mental demand}, \textbf{effort}, and \textbf{frustration} show similar patterns. In these three subscales, scores were similar between both shelf layouts. \link and \col are followed by \outline and \arrow. While \link and \col have a similarly low score for \textbf{effort}, \link slightly outperforms \col for \textbf{mental demand} and \textbf{frustration}.  

Overall, the results resemble those of the category ratings: \col and \link performed the best, followed by \outline, and lastly \arrow. However, they are much less pronounced in NASA-TLX. 

\vsubsection{Qualitative analysis and comments}
As mentioned in \autoref{sec:methods}, we also included free-text forms for further feedback. In addition to mentioning strategies, there were comments on our task design, as well as on brushing and filtering methods. As this work focuses on highlighting and linking, we focus on feedback on these issues.

\textbf{Tasks can have---but do not need---strategies.}
Regarding strategies, one participant mentioned that they first \textit{"[read the task] very quickly while selecting the [components]"} (P6). For a later question, they clarified that they \textit{"expanded [their] previous strategy with the red stripes on the axes"} (P6). Another participant (P16) described a similar strategy for the statement task, since they also \textit{"[looked] at named shelves before tending to the task"} (P16). For the multi-selection task, one participant stated that they \textit{"[selected] the products on the shelves going from left to right"} (P16). Another participant claimed that they adapted their strategy \textit{"[to select] one object at a time instead of selecting all at once"} (P3) for the multi-selection and statement task, after trying out the \arrow condition.  

\textbf{\col and \outline require more movement.} A quarter of our participants said that it was easier to solve tasks with \col, and they were able to \textit{"[select] multiple products at once"} (P3) with \col and \outline. However, both techniques subjectively felt to involve more movement, since there is no \textit{"indication for objects outside the field of view"} (P12), and, hence, \textit{"[they] had to look at the shelves to find them, and had to look twice to make sure that [they] had not forgotten any item"} (P1).

\textbf{\outline is harder to perceive than \col.} Some participants argued that \textit{"for \outline, [they had] to pay more attention compared to \col because its colorful edge has a small size"} (P7) and \textit{"[it requires] more mental processing and visual search"} (P18). Furthermore, the participants argued that some products had similar boundaries, and, hence, \outline was harder to perceive as the contrast was \textit{"not always optimal"} (P17). 

\textbf{\link works well for out-of-view but worse in peripheral regions.} According to several participants, \link makes multi-selection tasks easier as they can \textit{"just follow the line [...] to select the item"} (P1) and \textit{"needed to pay less attention to the shelves directly"} (P18). \link is also effective for out-of-view targets as \textit{"the lines [helped] to indicate that the highlighted products are behind [you] without the need to turn around"} (P20). However, some participants criticized the visibility of \link in the peripheral regions. According to P2, these were ambiguous, and P12 said if the tablet hides some lines, they may be missed.

\textbf{\arrow requires patience and strategies for many targets.} Although P1 was able to follow the path of \arrow, most of the participants criticized that \arrow is \textit{"annoying"} (P7) and \textit{"exhausting for the eyes"} (P13). Although the general direction can be guessed, the movement of the arrow \textit{"[seemed] to be influenced by user orientation"} (P17), and, hence, it is difficult to find the actual target. According to several participants, \textit{"[they had] to wait to see where the arrows go, for each product separately"} (P5) and \textit{"if [they missed] the exact moment it hits, [they] had to wait again for the next one"} (P14). Therefore, they tried to \textit{"keep fully still"} (P7). Several participants also argued that they performed strategies for multi-selection and statement response tasks, as \arrow produced a \textit{"very crowded visual field"} (P18). For example, participants \textit{"tried to crouch to see more clearly if the arrows were flying only upward"} (P17) or \textit{"searched for an outlier"} (P18).

%-------------------------------------------------------------------------
\vsection{Discussion}
%%%%%%%%%%%%%%%%%%%%%%%%%%%%%%%%%%%%%%%%%%%%%%%%%%%%%%%%%%%%%%%%%%%%%%%%%%%
\label{sec:final}
We conclude our paper with a summary of our observations, implications for future work, limitations, and some final remarks. \autoref{sec:discussions} discusses the results based on the guiding questions.

\vsubsection{Observations and lessons learned}
\label{sec:discussions}
\paragraph{\col and \link perform the best in terms of completion time, followed by \outline and \arrow.}
Our results show that \col and \link resulted in the lowest completion time. \col was slightly faster for the Outside-FOV, whereas \link was slightly faster for the Inside-FOV. While we expected \link to perform well as per GQ1, \col was surprisingly able to handle the Outside-FOV condition despite its lack of support for out-of-view and explicit link representations. We assume that this is due to the strong contrast of \col, which makes it a valid choice when a target search area or direction is known in advance. \link may have underperformed for Outside-FOV as it would pass directly through the participant's own body, making it difficult to visually trace it. Although \outline highlights a target in a similar manner as \col, it has similar mean completion time for the Inside-FOV, but not for the Outside-FOV. According to multiple participants (P2, P5, P7), with a more solid and better contrast of \outline, it would have similar benefits as \col. For example, P2 argued that \textit{"[\outline was] much harder to see [...] than the products that were highlighted [by \col{}]"} (P2). Lastly, \arrow was clearly the slowest as expected in GQ1. Participants clearly summarized the issue: \textit{"for [\arrow{}] I had to wait to be sure where it goes"} (P5), while other techniques such as \col and \link \textit{"made it possible to solve problems at a glance"} (P18). This aligns with known reasoning for why static representations can be superior to animations \cite{heerAnimatedTransitionsStatistical2007,tverskyAnimationCanIt2002}. It is possible that a combination of static highlights with salient animations could be effective, though at the cost of increased visual clutter.

\vparagraph{\col had the least amount of errors, with \outline, \link, and \arrow performing increasingly worse.}
We found that \col was clearly the most accurate highlighting technique, which once again may be due to its strong contrast. Note that, in several instances, errors were caused by mistakes in brushing the scatterplot or selecting the right product---likely caused by the challenging raycast interactions. However, as these issues affect all highlighting techniques uniformly, we believe that this does not invalidate our results. Therefore, the lack of errors of less than 1 m for both \col and \outline indicates that, when the product was correctly brushed, they resulted in little to no mistakes. In contrast, \link and \arrow performed much worse for errors, probably because they only indicate the center point of the target. As they do not indicate the contours of the target, like \col or \outline do, participants instead selected similar-looking nearby variants (e.g., different yogurt flavors). Therefore, we argue that, when a task involves similar varieties or distractors, \col and \outline can reduce ambiguities over \link and \arrow---especially if it is unclear if the referents stand alone individually or are grouped as one.

\vparagraph{\col and \link are subjectively preferred.}
We found that \col was generally preferred over \link. However, \col and \link differ significantly from \outline and \arrow. Similar results can be seen for enjoyment. We argue for GQ2 that both \col and \link are preferred over the remaining techniques. Given that \col and \link address different properties for a highlighting technique in situated analytics, it seems that the subjective preference for either of them would depend on the use case. A distinct advantage of \col is that it is easy to understand and has a bright color. One participant stated that \textit{"[\col{}] seemed the easiest [technique] to me"} (P20). Unlike \col, \link supports finding targets which are out of view, as one can solely follow the line from its beginning to its end. Several participants mentioned that \link highlighted targets behind them well (P12, P17, P18). However, in the periphery, it can be difficult to accurately determine the end point of a link (P2). Hence, we speculate that increasing the link thickness with increasing distance or using a combination of \col and \link may increase overall satisfaction, especially for real-world awareness. 

\vparagraph{\arrow is the least preferred despite its ability to direct to out-of-view targets.}
Since \arrow clearly has the lowest results for general experience and enjoyment, we can declare it the least preferred technique for GQ2 and GQ3, at least, for our scenario. This observation is also reflected in the comments of our participants, who described \arrow as \textit{"irritating"} (P1), \textit{"annoying"} (P7) or even \textit{"exhausting for the eyes"} (P13). Gruenefeld et al.~\cite{gruenefeldFlyingARrowPointingOutofView2018} had similar task performance results, but found a lower workload for \arrow, which could be due to our scenario with its many targets. Our animated \arrow resulted in worse results than the static arrows of Wieland et al.~\cite{WielandJ2022} and Yu et al.~\cite{YuD2020}, which could be due to its usage as a baseline condition based on familiarity reasons. Since Wieland et al.~\cite{WielandJ2022} only considered a single target at a time, we assume that their arrow approach would also result in clutter for multiple targets. One reason might be the clutter introduced by \arrow that affects its ability to highlight many targets. P12 observed that \textit{"[\arrow is] obstructing too much of the view, especially when there are multiple targets"} (P12). However, \link was not perceived as cluttered, suggesting that animation might causes poor scalability. Additionally, all previous approaches \cite{gruenefeldFlyingARrowPointingOutofView2018,WielandJ2022,YuD2020} included estimation tasks that could have led to better results for \arrow if applied in our context. One could modify \arrow such that it is not affected by head movements if many targets are present. However, the clutter of \arrow remains a major drawback.

\vparagraph{\outline is the most subtle and least obstructing.}
While \outline has the lowest mean rating in terms of subtlety, all methods except \arrow rank similarly in terms of obstruction. However, comments from participants, such as \textit{"I had to look at each product to be really sure it is not highlighted"} (P5), suggest that \outline is indeed the most subtle regarding our GQ4. We assume that the contours of \outline are perceived as part of the real scene and not as artificial highlighting. Although we expected \link to be subtle as well, the end of a line appears more artificial. One reason for the results of \arrow may be the animations, which potentially obstruct the field of view. While subtlety may not be a virtue for fast visual search or classical brushing and linking applications, there are scenarios where a subtle, non-obstructive technique with high real-world is beneficial. For example, we want to avoid walking into an object or another person while moving toward the selected referent.

\vparagraph{Brushing in situated analytics matters.} 
We also received several comments on the brushing component of our prototype. Participants agreed that it was easy to brush, as the \textit{"red marked areas on the axis [helped] a lot"} (P12), indicating that their cognitive load was reduced as intended. Participants however still struggled with adjusting the filter handles using the raycast interactions, which \textit{"affected [their] frustration"} (P17).  P16 also wanted \textit{"to see how many [data marks] are selected on the tablet"}, which would be needed in a proper SitA application. Any such improvements to brushing would naturally benefit any immersive application, not just SitA.

\vsubsection{Example scenarios and applications}
Based on our results, we see the following practical usage for our techniques. \col is valid to use for targets behind the user due to its bright color. However, the scenarios should not require the perception of the actual appearance, but only the silhouette of the target. We further assume that the general direction of the target must be known in advance. Unlike the more obtrusive \col, \outline is perceived as more subtle, but harder to see. However, one can see the actual referent's appearance. For example, guiding a trainee to a desired tool during service. In contrast to \col and \outline, \link might be a better choice for targets that have a greater distance and are potentially occluded by other objects. We suspect \link to be a valid choice in wide search spaces such as a library in which books are distributed across many shelves and floors. Based on the results, \arrow required more work to find the correct target. We assume that scenarios which primarily require direction guidance benefit from a limited number of arrows.

\vsubsection{Future directions} 
We intentionally used baseline highlighting techniques established in brushing and linking and attention guidance. Based on our results, there are clear avenues for improving each of the four techniques. 
Although we did not directly evaluate the matter, \col lacks a way to perceive the actual target (since it is distorted by the color). A modified \outline with increased contrast and outline width could inherit most of the benefits of \col, while not (fully) occluding the underlying targets.  For \link, we suggest that adjusting size based on distance, as well as adding a way of perceiving lines behind the tablet or user, would further increase the already encouraging results. However, care has to be taken to avoid excessive visual clutter~\cite{Hurter2019}. We assume that \arrow might be more beneficial if the user's movements did not have an impact on its path, or if used as a supplementary indicator combined with \col or \outline. 
In general, it seems likely that a hybrid technique can overcome the weaknesses of individual techniques, provided that we can avoid too much clutter. However, the optimal choice will likely depend on the scenario. For example, applying \col to large objects can be visually overwhelming, even if \col was the most favored overall. Future work may seek to devise an adaptive highlighting technique to tune the visual characteristics of the highlighting to the current viewing situation. 

Future work should also investigate brushing. If brushing in \textit{reverse}, i.e., brushing physical referents, an analyst could compare the values of nearby referent after physically navigating to an interesting real-world area, such as inspecting the bargain bin in the supermarket. Several techniques could support brushing in this manner, including spatial filtering~\cite{Feiner1993, Julier2000}, aggregation~\cite{Tatzgern2016}, and selection~\cite{sidenmarkOutlinePursuitsGazeassisted2020} techniques. We leave the exploration of this inverse brushing for future work.

While we believe the \supermarket{} is generalizable to other scenarios, our study lacked several conditions that should be explored in future work. First, our shelf layouts did not force participants to move because of occlusions. We assume out-of-view highlighting techniques such as \link are more beneficial here, especially, since they can also help with wayfinding.
%In addition, we assume that our shelves can easily be adapted to other scenarios like bookshelves in a library or different rooms in a building.
Second, we chose only static referents, whereas referents can also be dynamic (e.g., autonomous robots, animals). Third, our referents are at a comfortable human scale when compared to buildings or landscapes. These characteristics not present in our scenario require further evaluation.

\vsubsection{Limitations}
Some limitations can potentially affect the interpretation of the findings reported in our work. First, our small sample size and the chosen tasks could affect the results and design considerations of our work. In addition, we adjusted and calculated our linking completion time based on the last selection on the tablet, assuming that participants would first brush all data marks and then select the respective targets on the shelves. While we manually adjusted the timestamps for participants who obviously used a different strategy, it is possible that some participants missed a data mark during the multi-selection task. Hence, some timestamps may be non-representative. Since we used an open experimenter study design, we assume that this rarely happened and does not limit our results.

Another limitation is our use of VR to simulate AR, even if transferability from VR to AR has been demonstrated for other tasks \cite{21LeeC2009Replication, 23LeeC2013Replication}.
We assume that any differences observed between the highlighting techniques in our study will still translate into AR, and the reasons for using one technique over the other remain the same. However, it is likely that their effectiveness in AR will largely be dictated by the technical capabilities of the AR headset and tracking solution~\cite{Sousa2023}. For example, tracking drift may cause noticeable misalignment in object silhouettes when using \col, which may justify the use of other techniques less sensitive to this issue like \link or \arrow. The study of these considerations lies firmly outside the scope of this work.

%-------------------------------------------------------------------------
\vsection{Conclusions}
%%%%%%%%%%%%%%%%%%%%%%%%%%%%%%%%%%%%%%%%%%%%%%%%%%%%%%%%%%%%%%%%%%%%%%%%%%%
\label{sec:conclusions}
We present an initial exploration of brushing and linking for situated analytics in AR. We describe four highlighting techniques, representing conventional approaches to brushing and linking and attention guidance. We used these techniques in an exploratory user study and found that \col is still a reasonable choice as a highlighting technique for situated brushing and linking. We also saw that \arrow, a highlighting technique that requires spatial interpretation, performed worst in terms of both task performance and user experience. Based on our overall results, we propose that a combination of highlighting techniques might reap the benefits while overcoming individual weaknesses.
We consider our work a starting point to bring brushing and linking into the real and merging it with previous techniques designed for ``reading'' the environment.

\paragraph{Acknowledgments.}
We thank DFG (495135767 and EXC 2120/1–390831618), Carl-Zeiss-Stiftung (P2016-03-004), FWF (I5912-N), and the Alexander von Humboldt Foundation funded by the German Federal Ministry of Education and Research.

%-------------------------------------------------------------------------
% bibtex
\bibliographystyle{eg-alpha-doi}  
\bibliography{egbibsample}        

% biblatex with biber
%auskommentiert
%\printbibliography                

\end{document}